\documentclass[prd, aps, superscriptaddress, preprintnumbers, twocolumn, floatfix, nofootinbib]{revtex4-1}

\usepackage{amsfonts}
\usepackage{amsmath}
\usepackage{amssymb}
\usepackage{bm}
\usepackage{dcolumn}
\usepackage{epsfig}
\usepackage{graphicx}   
\usepackage{graphics}
\usepackage[latin1]{inputenc}
\usepackage{latexsym}
\usepackage{rotating}
\usepackage{hyperref}

\usepackage{amsfonts}
\usepackage{amsmath}
\usepackage{amssymb}
\usepackage{bm}
\usepackage{dcolumn}
\usepackage{epsfig}
\usepackage{graphicx}
\usepackage{graphics}
\usepackage[latin1]{inputenc}
\usepackage{latexsym}
\usepackage{rotating}
\usepackage{hyperref}
\usepackage{color}
\usepackage{soul}
\usepackage[normalem]{ulem}
\newcommand\be{\begin{equation}}
\newcommand\ba{\begin{eqnarray}}
\newcommand\ee{\end{equation}}
\newcommand\ea{\end{eqnarray}}

\begin{document}

\title{Constraining the Break of Spatial Diffeomorphism Invariance with Planck Data} %/ CMB Constraints on a Blue Tensor Tilt from Diffeomorphism Invariance Breaking}

\author{L. L. Graef} {\email{leilagraef@on.br}
%\affiliation{Observat\'orio Nacional, Rua General Jos\'e Cristino 77, S\~ao Crist\'ov\~ao, 20921-400,Rio de Janeiro, RJ, Brazil}

\author{M. Benetti} {\email{micolbenetti@on.br}
%\affiliation{Observat\'orio Nacional, Rua General Jos\'e Cristino 77, S\~ao Crist\'ov\~ao, 20921-400,Rio de Janeiro, RJ, Brazil}

\author{J. S. Alcaniz}{\email{alcaniz@on.br}

\affiliation{Departamento de Astronomia, Observat\'orio Nacional, 20921-400, Rio de Janeiro, RJ, Brasil}

\date{\today}

%%%%%%%%%%%%%%%%%%%%%%%%%%%%%%%%%%%%%%%%%%%%%%%%%%%%%%%%%%%%%%%%%%%%%%%%%%%%%%%%%%%%%%%%%%%%%%
\begin{abstract}

The current most accepted paradigm for the early universe cosmology, the inflationary scenario, shows a good agreement with the recent Cosmic Microwave Background (CMB) and polarization data.  However, when the inflation consistency relation is relaxed, these observational data  exclude a larger range of red tensor tilt values, prevailing the blue ones which are not predicted by the minimal inflationary models. Recently, it has been shown that the assumption of  spatial diffeomorphism  invariance breaking (SDB) in the context of an effective field theory of inflation leads to interesting observational consequences. Among them, the possibility of generating a blue tensor spectrum, which can recover the specific consistency relation of the String Gas Cosmology, for a certain choice of parameters. We use the most recent CMB data to constrain the SDB model and test its observational viability  through a Bayesian analysis assuming as reference 
%the $\Lambda$CDM+$r$ model
an extended $\Lambda$CDM+tensor perturbation model, which considers a power-law tensor spectrum 
%with the amplitude \textcolor{red}
parametrized in terms of the tensor-to-scalar ratio, $r$, and the tensor spectral index, $n_t$.
If the inflation consistency relation is imposed, $r=-8 n_t$, 
we obtain a \textit{strong} evidence in favor of the reference model whereas if such relation is relaxed, % ($n_t$ is left free to vary), 
 a \textit{weak} evidence in favor of the model with diffeomorphism breaking is found. We also use the same CMB data set to make an observational comparison between the SDB model, standard inflation and String Gas Cosmology.

\end{abstract}
%%%%%%%%%%%%%%%%%%%%%%%%%%%%%%%%%%%%%%%%%%%%%%%%%%%%%%%%%%%%%%%%%%%%%%%%%%%%%%%%%%%%%%%%%%%%%%

\maketitle

%%%%%%%%%%%%%%%%%%%%%%%%%%%%%%%%%%%%%%%%%%%%%%%%%%%%%%%%%%%%%%%%%%%%%%%%%%%%%%%%%%%%%%%%%%%%%%
\section{Introduction} 

The large amount of observational data recently released \cite{Hinshaw:2012aka, Aghanim:2015xee, BICEP2, Array:2015xqh, Ligo, Ligo2, Virgo, Virgo2} have allowed to test the viability of a wide range of models of the early universe. Much more precise observations are expected to become available in the next years, and much of the future efforts will be focused on the tensor data \cite{pixie,litebird,core,advligo,advact,actpol, PT}. The current constraints on the cosmological parameters are  compatible with both the simplest slow-roll scenarios of inflation \cite{infl} as well as with some alternative scenarios \cite{alternativas, alternativas2, alternativas3}. The prospects for improvement of these constraints with the next generation of experiments suggest us to keep investigating different possibilities. 
In particular, a possible detection of a blue tensor tilt in the future would rule out a large class of models of inflation, since the so-called consistency relation would not be fulfilled  in the standard scenario
without violating the Null Energy Condition (NEC).
It is worth mentioning that the recent polarization data have excluded a much larger range of red tensor tilt values than of blue ones \cite{Planck2015, it}.

In the inflationary context, the violation of the NEC without incurring in instabilities was achieved in some class of models~\cite{G_infl, Ghost_infl}.  At the same time, models with production of particle during inflation \cite{berera-rhb, Benetti:2016jhf}, axion inflation \cite{119, 120} or model with higher-curvature corrections to the gravitational effective action \cite{121, 122} are able to change the consistency relation, allowing a blue tensor tilt.  It was recently suggested that by breaking the assumption of spatial diffeomorphism invariance in the context of the Effective Field Theory (EFT) of inflation, a blue tensor spectrum can be achieved without violating the NEC \cite{graef, wands, nongauss, lin}. In this Spatial Diffeomorphism Breaking (SDB) model, the temporal diffeomorphism invariance is broken due to the time-dependence of the background. However, in the high energy regime of the early universe it is also possible that  the spatial diffeomorphism invariance is broken (we refer the reader to \cite{wands} for a detailed discussion)}.

The EFT approach corresponds to the description of a system through the lowest dimension operators compatible with the underlying symmetries.  This approach allows to characterise all the possible high energy corrections to simple inflationary scenario, whose sizes are constrained by experiment. It also has the advantage of being able to describe in a common language all single field models of inflation using only symmetry principles. This theory has been applied in recent works to describe the theory of fluctuations around an inflating cosmological background \cite{18}, and in this context was built the model considered here. In Ref.~\cite{graef} it was shown that the the scalar power spectrum of the EFT model with SDB exhibits  characteristic signatures which allow it to be easily confronted with the data. In what concerns the tensor spectrum, Ref.~\cite{wands} showed that the extra terms that arise from the SDB invariance can produce a blue tilt, $n_{T} > 0$, for a certain range of values of the parameter. 

Another viable model that predicts a blue tensor spectrum is the String Gas Cosmology (SGC) \cite{SG8, SG9, SG10},  an alternative scenario for the early universe that arises from a combination of basic principles of superstring theory and cosmology. In the SGC model of Refs.~\cite{rhb1, rhb2}, a consistency relation between the scalar spectral index $n_s$  (which corresponds to a red spectrum) and that of the tensor modes was derived, i.e.,
\be \label{sgcr}
n_{T}\approx -(n_{s}-1)\;.
\ee
On the other hand, current observations indicate that $n_s = 0.9655 \pm 0.0062$ (1$\sigma$) \cite{Planck2015}. Therefore, if the ratio $r$ of the tensor spectrum
to the scalar spectrum is not too small, then future CMB polarisation experiments may have the potential to differentiate between the simple inflationary consistency relation and the one from SGC. 

The consistency relation above can be reproduced by an EFT of inflation with breaking of spatial diffeomorphism invariance for a certain combination of the parameters~\cite{graef}. In order to distinguish these models with future observations it is important to analyse if such a  combination of parameters is allowed by the current data. This is one of the goals of the present work. By performing a Bayesian analysis using the most recent Planck and BICEP data, we test the observational  viability of SDB model in the context of the EFT of inflation. We also   discuss a possible observational distinction between the SDB model and the String Gas Cosmology.  This paper is organized as follows. Section~\ref{Sec:Spatial Diffeomorphism Breaking in the EFT} reviews the basic assumptions of the SDB model. In Section~\ref{Sec:Theoretical Analysis} we evaluate the power spectrum predicted by this model. 
The method, observational data sets and priors used in the Bayesian analysis are discussed in Section~\ref{Sec:Observational Analysis}. In Section~\ref{Sec:Results} we discuss our main results. We summarize our conclusions in Section~\ref{Sec:Conclusions}.

\section{Spatial Diffeomorphism Breaking in the EFT}
\label{Sec:Spatial Diffeomorphism Breaking in the EFT}

In what follows, we will consider the effective field theory for cosmological perturbations around a quasi-de Sitter background, with both  temporal and spatial diffeomorphism invariance broken (but with isotropy preserved).  We consider the following perturbed metric
\begin{equation}
ds^{2} \, = \, g_{\mu\nu}(x, \eta) dx^{\mu} dx^{\nu}\;.
\end{equation}
The metric fluctuations are defined by
\be
h_{\mu\nu}(x, \eta) \, = \, g_{\mu\nu}(x, \eta) - \bar{g}_{\mu\nu}(\eta)\;,
\ee
where $\eta$ is the conformal time and $\bar{g}_{\mu\nu}(\eta)$ is the background metric. We will focus on operators at most quadratic
in the fluctuations for simplicity. The approach usually adopted in the effective theory considers the unitary gauge, in which there are no perturbations of the inflaton field and all the fluctuations degrees are in the metric. 

Following Ref.~\cite{graef, wands, nongauss} the SDB invariance will be described through effective mass terms in the action for cosmological perturbations. These terms simply correspond to the most general way to express quadratic non-derivative operators in the  fluctuations that break the spatial diffeomorphism symmetry. To the  usual Einstein-Hilbert action expanded to second order, 
we add generic operators with no derivatives that are quadratic in the metric fluctuations $h_{\mu\nu}$. We consider the following action
\begin{align} \label{pdf1}
& S= \int d^{4}x \sqrt{-g} M_{pl}^{2} \left[R - 2\Lambda -2cg^{00}\right] \nonumber\\
& +\frac{1}{4} M_{pl}^{2} \int d^{4}x \sqrt{-g} [m_{0}^{2} h_{00}^{2} + 2m_{1}^{2} h_{0i}^{2} - m_{2}^{2} h_{ij}^{2} \nonumber\\ 
& + m_{3}^{2} h_{ii}^{2} - 2m_{4}^{2} h_{00} h_{ii}]. 
\end{align}
The terms in the first line are the ones which contribute to the homogeneous and isotropic background. The parameters $c$
and $\Lambda$ can be expressed as functions of the Hubble parameter ($H$) and its time derivative. The term proportional to  $m_{0}^{2}$ breaks time diffeomorphism invariance and the other mass terms break spatial diffeomorphism invariance.  The invariance is restored for the limit $m_{i} \rightarrow 0$ with $i \neq 0$. 

One may consider the mass terms in the above equation as arising from couplings between the metric and fields with a time-dependent profile during inflation. Following \cite{graef, wands}, we assume as an approximation that their coefficients are effectively constant during inflation, and go to zero after inflation ends.  However a small time-dependence proportional to the slow-roll parameter ($\epsilon$)
should be expected for these coefficients, which can be neglected for our purposes here.

As in \cite{graef}, we can define a particular combination of masses parameters,
\begin{equation} \label{pdfM}
M^{2} \, = \, \frac{-2m_{2}^{2} H^{2}(m_{2}^{2}-3m_{3}^{2}+(3+\epsilon)m_{4}^{2})}{(m_{0}^{2}+2\epsilon H^{2})(m_{2}^{2}-m_{3}^{2})+ m_{4}^{4}} \, .
\end{equation}
As it will become clear latter, there is an important quantity which appears in the predictions of the model, which we define as
\be\label{alfadef} 
\alpha \, \equiv \, \sqrt{2\epsilon \frac{M^{2}}{H^{2}}+3\epsilon +\frac{M^{2}}{H^{2}} +9/4} \, .
\ee
As usual, we can decompose the fluctuations in terms of scalar, vector and tensor 
perturbations, i.e.,
\begin{align}
& h_{00} =\psi, \\
\nonumber & h_{0i} = u_{i} + \partial_{i} v,   %\; \;  \; \; \; \; \; \; \; \; \; \; \; \; \; \; \; \; \; \; \; \; \; \; \; \; \; \; 
\; \; {\rm{with}} \; \;   \partial_{i} u_{i} =0, \\
\nonumber & h_{ij} = \chi_{ij} + \partial_{(iSj)} + \partial_{i} \partial_{j} \sigma + \delta_{ij} \tau, \; \; {\rm{with}} \; \partial_{iSi} =\partial_{i} \chi_{ij}=0. 
\end{align}
From the tensor part of the action it was obtained in \cite{wands} the following tensor power spectrum
\begin{align} \label{pdftensorspectrum}
P_{T} &= \frac{2H^{2}}{\pi^{2} M_{pl}^{2} c_{T}} \left(\frac{k}{k_{*}}\right)^{n_{T}}, \\
\label{pdftensorspectrum2}
n_{T} &=- 2\epsilon + \frac{2}{3} \frac{m_{2}^{2}}{H^{2}} \left(1+ \frac{4}{3}\epsilon\right),
\end{align}
to first order in the slow-roll parameter. In the equation above  
$c_{T}=1$~\cite{graef, wands}  and $k_{*}$ denotes the $k$ pivot. We can see that the parameters $m_0$, $m_1$, $m_3$ and $m_4$ do not appear in the action for tensor perturbations.
We can also see that if $m_{2}^{2}/H^{2}$ is 
positive and sufficiently larger than the slow-roll parameter, then
we have a positive tensor spectral index. 

On the other hand, from the scalar part of the action the following power spectrum was obtained in Ref.~\cite{graef}:
\begin{equation} \label{pdfscalarspectrum}
P_{\mathcal{R}} = \frac{k^{3}|v|^{2}}{a^{2}N^{2}} = \frac{k^{3-2\alpha}}{8\pi^{2}}\left(\frac{c_{s}^{-2\alpha}}{2a^{2}N^{2}}\right)\left(\frac{|\eta|}{\sqrt{\alpha^{2} -\frac{1}{4}}}\right)^{1-2\alpha},
\end{equation}
where
\begin{equation} \label{pdfcs}
c_{s}^{2} \, = \, 2 \epsilon H^{2}\frac{(m_{2}^{2} - 
m_{3}^{2})}{(m_{0}^{2}+2\epsilon H^{2})(m_{2}^{2}-m_{3}^{2})+ m_{4}^{4}} \, ,
\end{equation}
and
\begin{equation} \label{pdfN}
N^{2} \, \equiv \, \left(\frac{M_{pl}^{2}}{H^{2}}\right) \frac{(m_{0}^{2} + 2 \epsilon H^{2})(m_{2}^{2} - m_{3}^{2})+m_{4}^{4}}{2(m_{2}^{2} - m_{3}^{2})} \, .
\end{equation}
Note that in the limit when the extra terms in the action go to zero, $c_s^2 \rightarrow 1$ and $N^{2} \rightarrow \epsilon$. Note also  that $N^{2}$ can be written as a function of $c_{s}^{2}$, as $N^{2}=M_{pl}^{2} \epsilon / c_{s}^{2}$. Since  in this model the curvature perturbation is not constant after Hubble radius crossing \cite{wands}, 
an explicit time-dependence appears in  Eq. (\ref{pdfscalarspectrum}). We need to evaluate
the result at the end of inflation ($\eta= \rm{const.}$, $a = \rm{const.}$).

%We can see that the scalar spectral index is given by
%%
%\begin{equation} \label{pdfns1}
%n_{s} - 1\, = \, 3 -2\alpha \, = \, 3 - 2\sqrt{2\epsilon M^{2}/H^{2} + 3\epsilon + M^{2}/H^{2} + 9/4} \, .
%\end{equation}

We can write a simplified expression for the spectral index by expanding
the square root in the expression \ref{alfadef} for $\alpha$ as follows,
\begin{align} \label{pdfns2}
n_{s} -1 \, &= \, 3 -2\alpha \, = \, 3 - 3\sqrt{\frac{8 \epsilon}{9} \frac{M^{2}}{H^{2}} + \frac{4}{3} \epsilon 
+ \frac{4}{9}\frac{M^{2}}{H^{2}} +1} \nonumber  \, \\ &\approx \, - 2\epsilon - \frac{2}{3}\frac{M^{2}}{H^{2}} \, .
\end{align}
Although the scalar spectrum must be calculated at the end of inflation, it is possible to show that the slow-roll parameter that appears in the above equations  corresponds to the one at the moment of Hubble radius crossing \cite{graef}. In general, quantities calculated at horizon crossing are  marked with $``*"$ but, for simplicity, we will omit such a symbol. For instance, $\epsilon_{*}$ will be denoted just by $\epsilon$. On the other hand, quantities calculated at the end of inflation will be indicated as such.

As shown in \cite{graef}, the approximated SGC consistency relation (\ref{sgcr}) can be recovered in this model if the parameters satisfy the following relation,
\begin{equation}
-2\epsilon + \frac{2}{3} \frac{m_{2}^{2}}{H^{2}}(1+2\epsilon) \, = \, +2\epsilon + \frac{2}{3} \frac{M^{2}}{H^{2}}(1+2\epsilon) \, ,
\end{equation}
which can be written as a function of the parameter $\alpha$ by the  approximated expression
\begin{equation}
 \frac{m_{2}^{2}}{H^{2}} \approx -\frac{9}{2}+3\epsilon + 3\alpha.
\end{equation}
In section IV we shall use current observational data  to obtain the allowed range of values for $M^{2}$ and $m_{2}^{2}$ (or equivalently for $\alpha$ and $m_{2}^{2}$). 

\section{Theoretical Analysis}
\label{Sec:Theoretical Analysis}

Let us  analyse the expression of the power spectrum (\ref{pdfscalarspectrum}). Since we suppose the SDB to be small (and $c_{s}^{2}\rightarrow 1 $ in the absence of this breaking), then in many cases of interest we can express $c_{s}$ as $c_{s}=1-\delta$, where $\delta$ is much smaller than 1. By expressing the quantity $N^{2}=M_{pl}^{2} \epsilon / c_{s}^{2}$  and considering $c_{s}^{2}\approx 1$, we can obtain a much simpler expression for the power spectrum,
\be \label{simple}
P_{\mathcal{R}} \, = \,  \frac{k^{3}|v|^{2}}{a^{2}N^{2}} = \frac{k^{3-2\alpha}}{8\pi^{2}}\left(\frac{1}{2a_{end}^{2}\epsilon}\right)\left(\frac{|\eta_{end}|}{\sqrt{\alpha^{2} -\frac{1}{4}}}\right)^{1-2\alpha},
\ee
which is a function of a single free parameter $\alpha$ encoding the effects of the SDB. Although this approximation for the amplitude only contemplates a class of the SDB models, it  almost does not affect the constraints we will derive on the spectral indexes, so that the results for these quantities remain quite general. We write the conformal time at the end of inflation $\eta_{end}$ and the scale factor at the end of inflation $a_{end}$ as a function of the respective quantities in the moment of Hubble radius exit and also as a function of the number of e-folds $\mathcal{N}$, i.e.,
\be
a_{end} \approx a_{*}e^{\mathcal{N}}, \, \,  \,  \, \, \, \, \, \, \,  \, \, \, \,  \, \eta_{end} \approx e^{-\mathcal{N}}\eta_{*} \approx \frac{-e^{-\mathcal{N}}}{H_{*}a_{*}}.
\ee
By substituting the above expressions in Eq. (\ref{simple}), and omitting the Hubble radius crossing index $*$, we obtain
\be \label{PN}
P_{\mathcal{R}} \, \approx \,  \frac{1}{8\pi^{2}}\left(\frac{k}{k_{*}}\right)^{3-2\alpha}\left(\frac{H^{2}}{2\epsilon e^{2\mathcal{N}}}\right)\left(\frac{e^{-\mathcal{N}}}{\sqrt{\alpha^{2} -\frac{1}{4}}}\right)^{1-2\alpha} ,
\ee

In the case in which spatial diffeomorphism is preserved, the curvature perturbation is conserved after Hubble radius crossing, and the power spectrum is then computed at this crossing moment. Therefore in order to recover the usual case we substitute $a_{end} \rightarrow a_{*}$ and $\eta_{end} \rightarrow \eta_{*}$, which can be done by making $\mathcal{N}=0$ in the above equation. Also, in this limit, we can see from Eq. (\ref{pdfns2}) that by making $M=0$ (i.e. neglecting the contribution from SDB), we obtain $\alpha \approx 3/2 +\epsilon$. We can see that by considering these values in the above equation we recover the usual expression for the power spectrum of inflation, as expected. 

For the purpose of making a theoretical analysis here let us write $\alpha$ as $\alpha \approx 3/2 +\delta_{\alpha}$, where $\delta_{\alpha}$ encompasses the contribution from $\epsilon$ plus the contribution coming from the SDB ($\delta_{\alpha} \approx \epsilon + M^{2}/(3H^{2})$ from Eq.~\ref{pdfns2})\footnote{$M^{2}$ can be either positive or negative, just like $m_{i}^{2}$.}. Using this notation in the above equation we obtain
\be \label{PN1}
P_{\mathcal{R}} \, \approx \,  \frac{1}{8\pi^{2}}\left(\frac{k}{k_{*}}\right)^{-2\delta_{\alpha}}\left(\frac{H^{2}}{\epsilon }\right)e^{2\mathcal{N}\delta_{\alpha}} \, ,
\ee
where the number of e-folds between the Hubble radius crossing and the end of inflation is usually considered to be $\mathcal{N} \approx 60$. Note that, if $\alpha=3/2$, and consequently $\delta_{a}=0$, a Harrison-Zeldovich spectrum is obtained. For $\delta_{a}=\epsilon$ we have an almost scale invariant spectrum with the small red tilt, as predicted by usual inflationary models. In this case, the amplitude of the power spectrum is given by $A_{s}\approx H^{2}/(8\pi^{2}\epsilon)$. We can see from the above equation that, 
since $\delta_{a}$ is in the exponent and it is multiplied by $\mathcal{N}$, the effect of the SDB  is exponentially amplified in the amplitude of the scalar power spectrum. Therefore, one must expect that the  combination of mass parameters described by $M^{2}$ will be quite constrained around zero. The same, however, is not expected for the parameter $m_{2}^{2}$, which enters in the expression of the tensor tilt. 

\section{Analysis Method}
\label{Sec:Observational Analysis}

In order to compute the CMB anisotropies spectrum for a given range of values of the $\alpha$, $H^2$ and $n_t$ parameters, we use a modified version of the  {\sc CAMB} code~\cite{camb} where the primordial power spectra are given by Eqs.~(\ref{pdftensorspectrum}) and~(\ref{PN}). We performed a Monte Carlo Markov chain analysis using the available package {\sc CosmoMC}~\cite{cosmomc}  and implementing the nested sampling algorithm of {\sc Multinest} code~\cite{mn1,mn2, mn3} to obtain our results and calculate the Bayesian evidence factor.  
In our Bayesian analysis we used the most accurate Importance Nested Sampling (INS)~\cite{Cameron:2013sm, mn3} instead of the vanilla Nested Sampling (NS), requiring a INS Global Log-Evidence error of $\leq 0.02$.
%%%%%%%%%%%%%%%%%%%%%%%%%%%%%%%%%%%%%%%%%
%models

In addition to the above primordial parameters, we vary the usual set of cosmological variables, namely the baryon density, %$\Omega_bh^2$, 
the cold dark matter density, 
%$\Omega_ch^2$, 
the ratio between the sound horizon and the angular diameter distance at decoupling, 
%$\theta$, 
and the optical depth:
%, $\tau$. 
%
%\begin{equation}
%\nonumber
$\left \{\Omega_bh^2~,~\Omega_ch^2~,~\theta~,~\tau \right \}$.
%\end{equation}
%
We consider purely adiabatic initial conditions, fix the sum of neutrino masses to $0.06~eV$, and also vary the nuisance foregrounds parameters~\cite{Aghanim:2015xee}.

It is important to emphasize that in SDB model the scalar primordial potential is not parameterized with the usual primordial scalar amplitude, $A_s$, and primordial spectral index $n_s$, as is done for the minimal $\Lambda$CDM model.
We consider as reference for our comparison two models, namely: the $\Lambda$CDM+$r$, where r is the tensor-to-scalar ratio parameter, which assumes inflation relation consistency ($r=-8 n_t$) and the $\Lambda$CDM+$r$+$n_t$, where the inflationary relation consistency is relaxed.

In our  analysis we work with the following priors on the SDB primordial parameters:
%%%%%%%%%%%%% TAB: priors %%%%%%%%%%%%%%%%%%%%%%%%%%
%
\begin{eqnarray}
\nonumber
-2< &\ln (10^{10}H^{2}) & < 3.5 ,\\
\nonumber
 1.48 < &\alpha & < 1.54 ,\\
 \nonumber
-1.5 < &n_{t} & < 5.5 .
\end{eqnarray}
We also assume that the $\epsilon$ and $\mathcal{N}$ values do not change significantly with respect to the canonical inflationary chaotic model, and fix them to $\epsilon=1/121$ and $\mathcal{N}=60$. The parameter $\alpha$ mimics the $\Lambda$CDM spectral index ($n_s$). Thus, from Eq. (\ref{PN}) one may expect values around $\alpha \simeq 1.51$. At the same time, joint effects of $H^2$ and $\alpha$ parameters replace the canonical primordial scalar amplitude. In such a degeneration of parameters, the amplitude increases considerably if a small increase in the value of $\alpha$  is made. Therefore, the $H^2$ value must compensate it by varying in a larger interval.

%data
We use the CMB data set from the latest Planck Collaboration release~\cite{Planck2015},  considering the high multipoles Planck temperature data (\textit{TT}) from the 100-,143-, and 217-GHz half-mission T maps, and  the low multipoles data (\textit{low-P}) by the joint TT, EE, BB and TE likelihood, where EE and BB are the E- and B-mode CMB polarization power spectrum and TE is the cross-correlation temperature-polarization. We also use B-mode polarization data from the BICEP2 Collaboration \cite{BICEP2, Array:2015xqh} to constraint the parameters associated to the tensor spectrum, using the combined BICEP2/Keck-Planck likelihood, hereafter \textit{BKP}. Finally, we perform our analysis for a scalar and tensor pivot scale $k_*=0.01$ $\rm{Mpc}^{-1}$, since at this value the BKP release data are most sensitive and it is close to the decorrelation scale between the tensor amplitude and slope for Planck and BKP joint constraints~\cite{Planck2015}.

In order to make an appropriate comparison between the SDB model and the $\Lambda$CDM scenario we use the Bayesian model comparison {{(see Refs.~\cite{Benetti:2017gvm, Benetti:2016ycg} for a more detailed discussion)}}. 
The Bayesian \textit{evidence} $\mathcal{E}$ is defined as the marginal likelihood for the model $M_i$:
\begin{equation}
\mathcal{E}_{M_i} =\int d\theta p(x|\theta, M_i) \pi(\theta | M_i)\;.
\end{equation}
where $x$ stands for the data, $\theta$ is the parameters vector and $\pi(\theta|M_i)$ the prior probability distribution function. The ratio of the Bayesian evidence of the two models (the so-called \textit{Bayes Factor}) can be defined as $\mathit{B}_{ij}= \mathcal{E}_{M_i}/\mathcal{E}_{M_j}$,
%
%\begin{equation}
%B= \frac{ \int d\theta' p(x|\theta' M') p(\theta'|M')}{ \int d\theta p(x|\theta M) p(\theta|M)}
%\mathit{B}_{ij}= \frac{\mathcal{E}_{M_i}}{\mathcal{E}_{M_j}}\;,
%\end{equation}
%
where $M_j$ is the reference model. We assume uniform  (hence separable) priors in each parameter, such that its possible to write $\pi(\theta|M) = (\Delta \theta_1 ~. . .~\Delta\theta_n)^{-1}$ and
\begin{equation}
\mathit{B}_{ij}= \frac{ \int d\theta p(x|\theta,M_i)}{ \int d\theta' p(x|\theta',M_j)}
\frac{(\Delta \theta_1 ~. . .~\Delta\theta_{n_i})}{(\Delta \theta'_1 ~. . .~\Delta\theta'_{n_j})}\;.
\label{eq:bayes}
\end{equation}

In order to classify the models  we use the Jeffreys scale \cite{Jeffreys} which gives empirically calibrated levels of significance for the strength of evidence. 
Here we will use a revisited version of the Jeffreys convention suggested in \cite{Trotta} where 
$ \ln \mathit{B}_{ij}  = 0 - 1 $,
$ \ln \mathit{B}_{ij}  = 1 - 2.5 $,
$ \ln \mathit{B}_{ij}  = 2.5 - 5 $,
and $ \ln\mathit{B}_{ij}  > 5 $ indicate an {\textit{inconclusive}}, {\textit{weak}}, {\textit{moderate}} and {\textit{strong}} preference of the model $M_i$ with respect to the model $M_j$.{\footnote{Note that $\ln \mathit{B_{ij}} < 0$  means support in favor of the reference model $M_j$.}} %(see ref.~\cite{Trotta, Santos:2016sti} for a more complete discussion about this scale).
%%%%%%%%%%%%%%%%%%%%%%%%%%
%%%%%%%%%%%%%%%%triangle_plot_cosmological%%%%%%%%%%%%%%%%%%%%%%
\begin{figure*}[!]
	\centering
	\includegraphics[width=0.6\hsize]{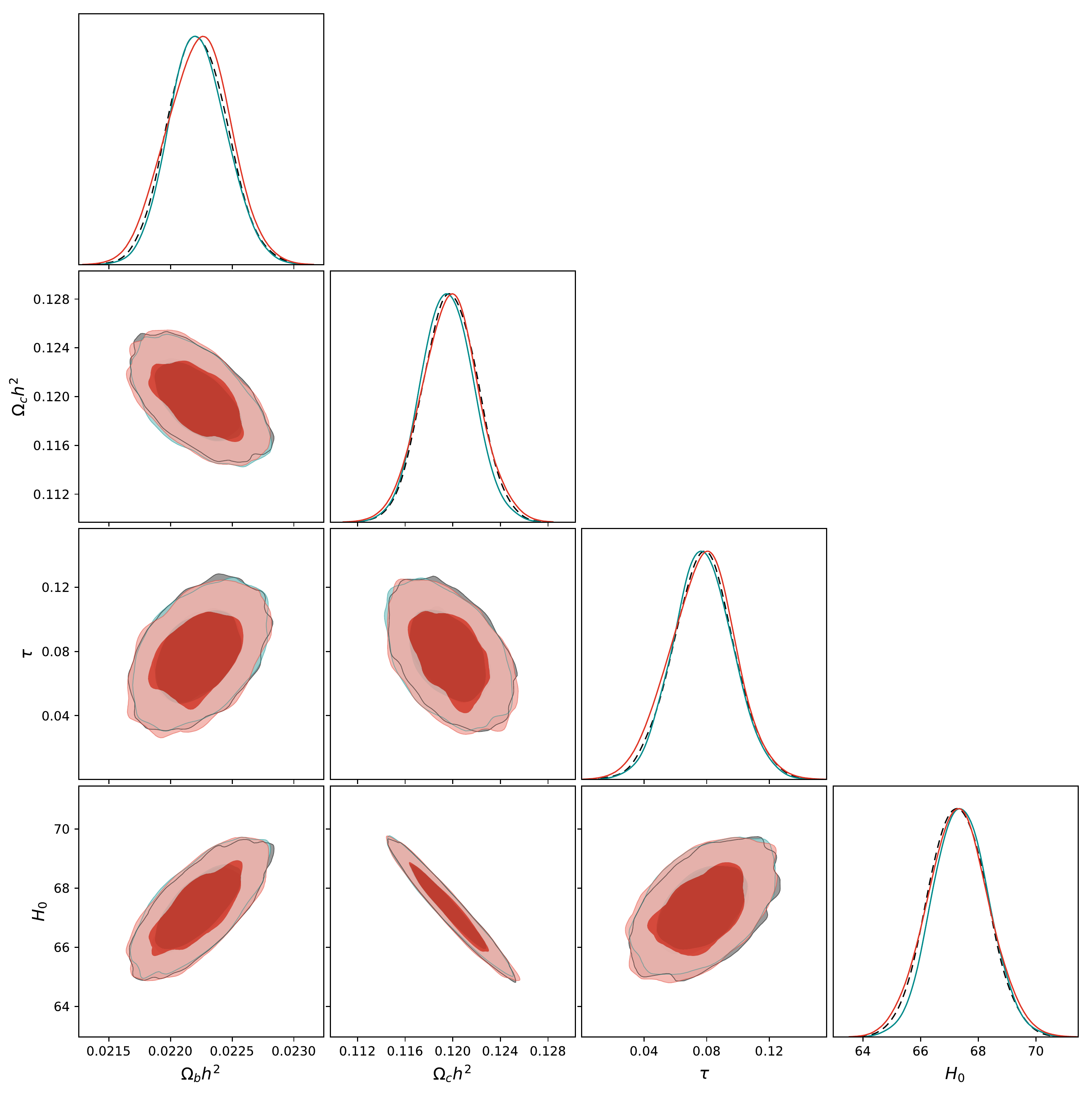}
	\caption{{The $68\%$ and $95\%$ confidence regions for the SDB (cyan), $\Lambda$CDM+$r$ (red) and $\Lambda$CDM+$r$+$n_{t}$ (black dashed) cosmological parameters  obtained from TT+lowP+BKP data.}
	\label{fig:tri_cosmological}}
\end{figure*}
%%%%%%%%%%%%%%%%%%%%%%%%%%%%%%%%%%%%%%
%
%%%%%%%%%%%%%%%%ns_logA/alpha_H2%%%%%%%%%%%%%%%%%%%%%%
\begin{figure}[!]
	\centering
	\includegraphics[width=1\hsize]{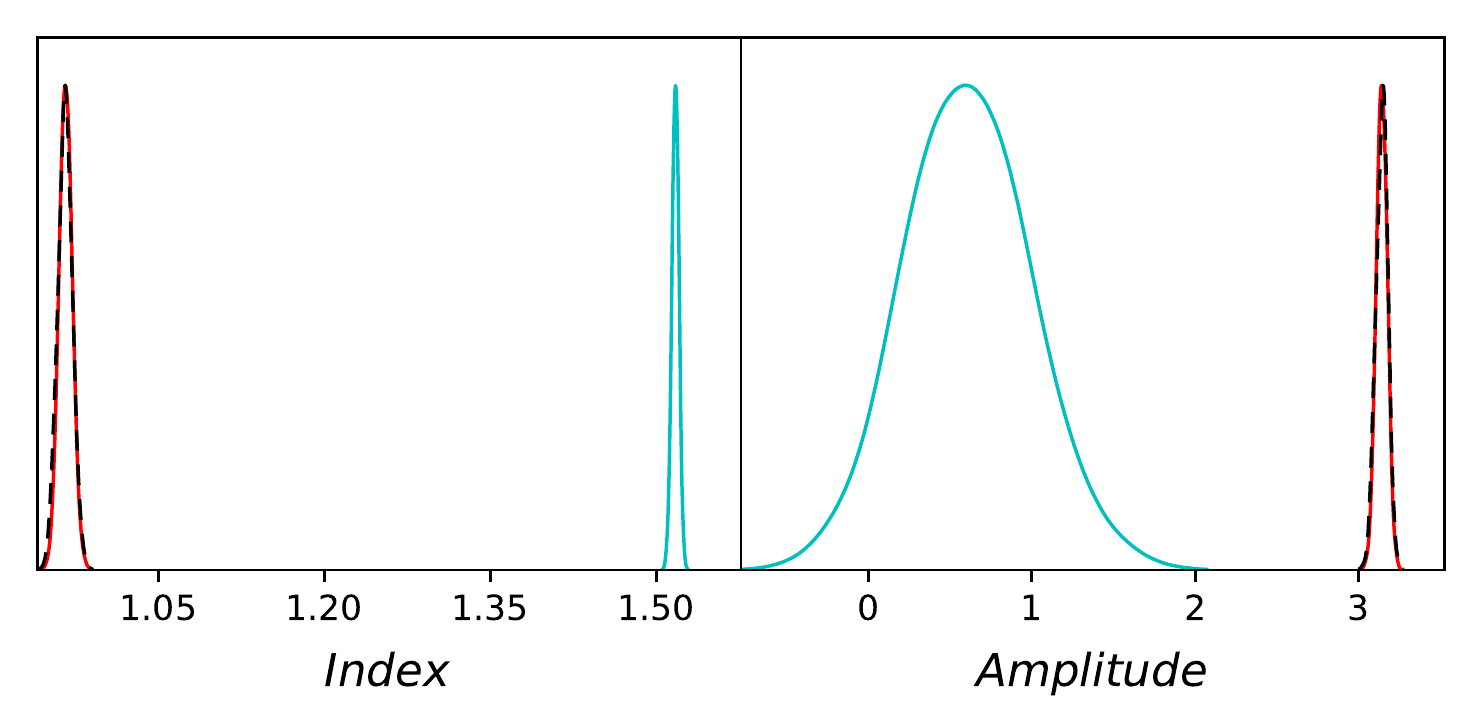}
	\caption{{Posterior of the index $\alpha$ and the amplitude $H^{2}$ parameters of the SDB model (cyan) compared with the index $n_s$ and the amplitude $A_s$ of the $\Lambda$CDM+$r$ model (red) and $\Lambda$CDM+$r$+$n_{t}$ model (black dashed). }
	\label{fig:ns_a_logA_H2}}
\end{figure}
%%%%%%%%%%%%%%%%%%%%%%%%%%%%%%%%%%%%%%

\section{Results}
\label{Sec:Results}

Fig.~(\ref{fig:tri_cosmological}) shows the good agreement between the cosmological parameters constrained in the context of the SDB model (cyan) and the reference models one, namely the $\Lambda$CDM+$r$ (red) and the $\Lambda$CDM+$r$+$n_t$ (black dashed). 
In the left panel of Fig.~\ref{fig:ns_a_logA_H2}, the posterior probability comparison between the scalar spectral index of the $\Lambda$CDM+tensor models and the $\alpha$ parameter of the SDB model shows a comparable goodness of constraint, while the shift of the mean value (as discussed in the previous section) depends on our parametrization choice. 
In the right panel of Fig.~\ref{fig:ns_a_logA_H2}, we show the posterior curves for the logarithmic values of the primordial scalar amplitude of the reference models and the $H^2$ parameter of the SDB. 
As expected, the degeneracy between the parameters $H^{2}$ and $\alpha$ (see Fig.~\ref{fig:H_alpha}) discussed in the previous section causes a worsening in the $H^{2}$ constraint. Our analysis show that the difference between the $H^{2}$ constraints and the primordial amplitude of is significant.

%comments on Fig. tri_tensorial
In Fig.~\ref{fig:posterior_tensorial}, we show the posterior density probability for the tensor spectral index and the tensor-to-scalar ratio parameters. 
The latter is treated as derived parameter in the SDB model and is calculated as the ratio between the tensor and the scalar power spectrum amplitude at the pivot scale $k_{0.01}$.
Firstly, we note that the data allow a larger range of blue tensor tilt, excluding negative values at 68\% C.L.
Such result is fully consistent with previous analysis from the Planck Collaboration~\cite{Planck2015}
 using both temperature and polarization CMB data, and with the
% where they used  TT,TE,EE+lowP for a $\Lambda$CDM+$r$+$n_t$ model (with relaxed inflation consistency relation). }
%On the other hand, when a joint BKP likelihood is considered, negative $n_{t}$ values can still be allowed in 1$\sigma$. In the 
recent analysis of Ref. \cite{it}, where joint CMB and BKP data are used.
%it is also  found that for the $\Lambda$CDM+$r$+$n_t$ model (using TT+lowP+BKP data) a red tensor tilt is excluded in $1\sigma$.
In the right panel of Fig.~\ref{fig:posterior_tensorial} we also note the very sharp constraints on $r_{0.01}$ for the SDB model.
%: at 1$\sigma$  $r_{0.01} = 0.017 \pm 0.007$, while $r_{0.01}<0.031$  within 95\% CL.
Such a narrow bound ($r_{0.01} = 0.017 \pm 0.007$ at 1$\sigma$) is due to our parametrization choice, since the tensor amplitude is directly linked to the scalar one (both depending on $H^2$). 
We stress that our parametrization worsens the constraints on the scalar amplitude but improves the tensor amplitude bound.

In Tab.~\ref{tab:Tabel_results_1} we summarize the results of our analysis for the $H^2$, $\alpha$ and $n_t$ parameters of the SDB model. The parameter $\alpha$ is constrained to be $1.517 \pm 0.003$ at 68$\%$ confidence level, which entails a corresponding $n_{s}$ value very similar to the standard inflationary prediction. This implies that the parameter $M^{2}/H^{2}$, which quantifies the effect of the diffeomorphism breaking in the scalar spectrum, is constrained to be very small. 
However, from Eq.~(\ref{pdfM}) we can see that it is possible to have $M \approx 0$ along with not so small values of $m_{2}$ ($m^{2}_{2}/H^{2}$ of order $\epsilon$, for example). 
%%%%%%%%%%%%%%%%%%%%%%%%%%%%%%%%%%%%%%%%%%%%%%%%%
\begin{table}
\centering
\begin{tabular}{|c|c|c|c|c}
\hline
{Parameter}&
{\textbf{$SDB \, \,  \, Model$}}\\
\hline
$\ln (10^{10}H^{2})$	
& $0.605 \pm 0.384$ 
\\
$\alpha$ 
& $1.517 \pm 0.003$ 
\\
$n_{t}$
& $1.914 \pm 1.150$
\\
\hline
\hline
$\Delta \chi^2_{\rm best}$         
%& $ 	  11305.2 -11305.2 $
& $ 0 $
\\
$\ln \mathit{B}_{ij}$ 
%& $ -5711.73 + 5695.96 $
& $ - 15.8$
\\
\hline
\hline
$\Delta \chi^2_{\rm best(+n_{t})}$         
%& $ - 11305.2 + 11309.13 $
& $ + 3.9 $
\\
$\ln \mathit{B}_{ij(+n_{t})}$ 
%& $ -5711.73 + 5713.36 $
& $ + 1.6$
\\
\hline
\end{tabular}
\caption{
$68\%$ confidence limits for the cosmological parameters  using TT+lowP+BKP data. 
The $\Delta \chi^2_{best}$ and the $\ln {B}_{ij}$ refers to the difference with respect to the $\Lambda$CDM+$r$ with fixed consistency relation. The subscript ``$_{(n_{t})}$" refers to the difference with respect to the $\Lambda$CDM+$r$+$n_t$ (with relaxed consistency relation).
\label{tab:Tabel_results_1}}
\end{table} 
%%%%%%%%%%%%%%%%%%%%%%%%%%%%%%%%%%%%%%%%%%%%%%%%%%
%

%
%%%%%%%%%%%%%%%%H^2_alpha%%%%%%%%%%%%%%%%%%%%%%
\begin{figure}[!]
	\centering
	\includegraphics[width=0.8\hsize]{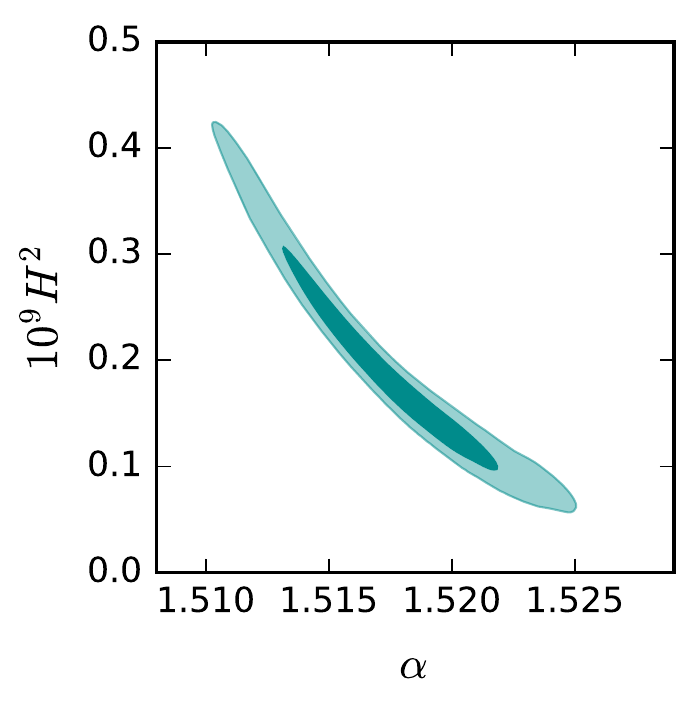}
	\caption{ {The $68\%$ and $95\%$ confidence regions for the parameters $H^{2}$ and $\alpha$. }
	\label{fig:H_alpha}}
\end{figure}
%%%%%%%%%%%%%%%%%%%%%%%%%%%%%%%%%%%%%%
%%%%%%%%%%%%%%%%triangle_plot_tensorial%%%%%%%%%%%%%%%%%%%%%%
\begin{figure}[!]
	\centering
	\includegraphics[width=1\hsize]{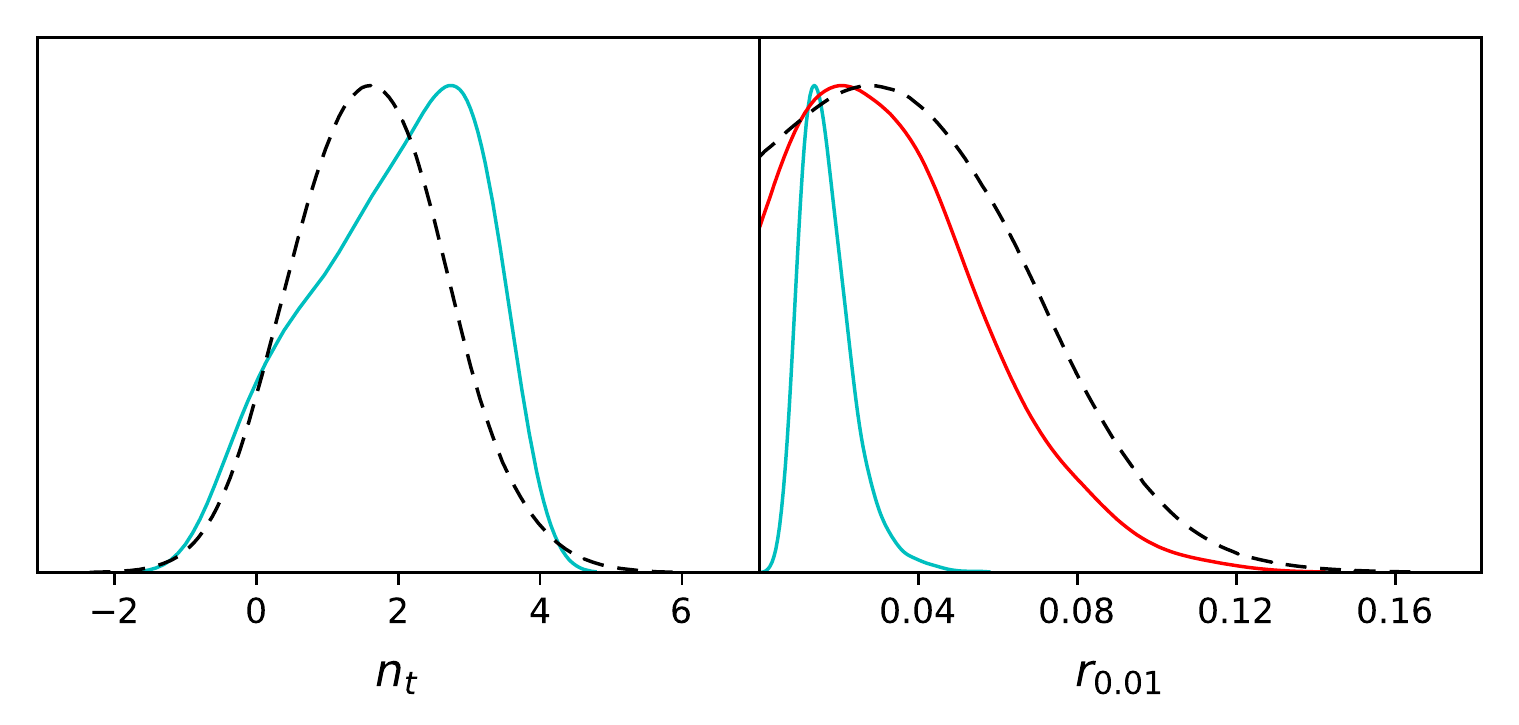}
	\caption{{Posterior density probability distribution for the SDB (cyan line) and $\Lambda$CDM+$r$ primordial tensor parameters with (red line) and without (black dashed line) inflation consistency relation arising from TT+lowP+BKP data.}
	\label{fig:posterior_tensorial}}
\end{figure}
%%%%%%%%%%%%%%%%%%%%%%%%%%%%%%%%%%%%%

We emphasize  that positive values of $n_{t}$ correspond to values of $m_{2}^ {2}$ such that $m_{2}^{2}/H^{2}>3\epsilon$ (see Eq. (\ref{pdftensorspectrum2})). A combined  analysis involving the scalar and tensor spectral indexes show that 
a red tensor tilt ($m_{2}^{2}/H^{2}<3\epsilon$) is only marginally allowed by the data at $1\sigma$, 
while a larger range of positive values for $m_{2}^{2}$  is supported {\footnote{We must point out that although the data prefer higher values of $m_{2}^{2}/H^ {2}$ this theoretical model is justified only for values of  $m_{2}^{2}/H^ {2} << 1$. However, for illustrative purposes, we show in the figures  all values for the tensor tilt allowed by the data. Nevertheless, the two lines shown in Fig.~\ref{fig:ns_nt}  are associated with values of $m^{2}_{2}$ lying within the range of strict validity of the theory.}}.

%
%%%%%%%%%%%%%%%nt_ns%%%%%%%%%%%%%%%%%%%%%%%%
\begin{figure*}[!]
	\includegraphics[width=0.3\hsize]{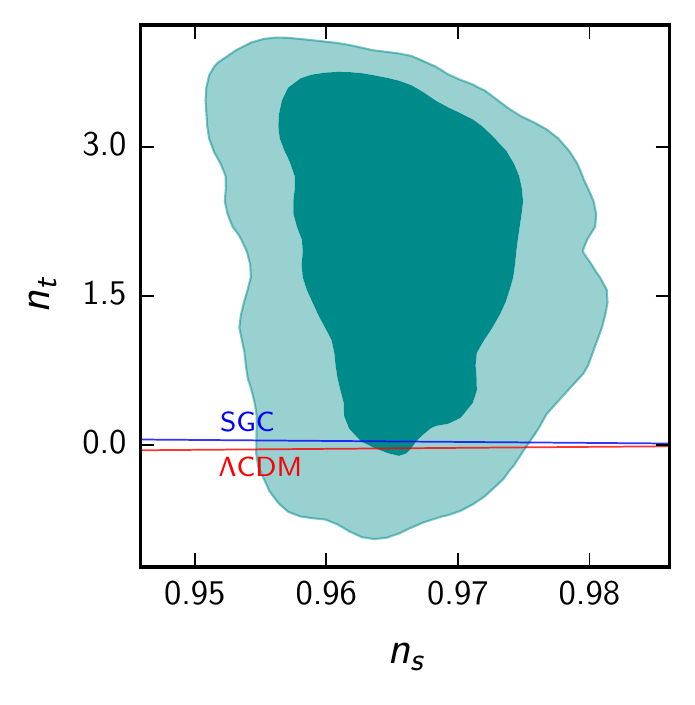}
		\includegraphics[width=0.28\hsize]
	{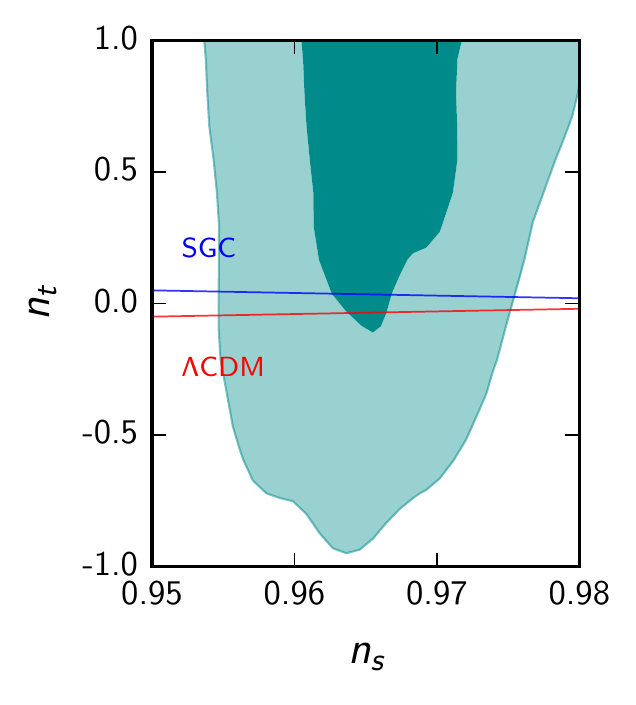}
	\caption{ {The $68\%$ and $95\%$ confidence regions for the parameters $n_{s}$ and $n_{t}$. The blue and the red lines correspond to the approximated consistency relations for SGC and standard inflation, respectively. 
The figure on the right corresponds to a zoom of the figure on the left.}
	\label{fig:ns_nt}}
	\label{fig:ns_nt_zoom}
\end{figure*}
%%%%%%%%%%%%%%%%%%%%%%%%%%%%%%%%%%%%%
%

In the last lines of the Table~(\ref{tab:Tabel_results_1}), we show the $\Delta\chi^2$ and the Bayes factor of such SDB model with respect to the $\Lambda$CDM+tensor scenarios, when the inflationary consistency relation is considered (first lines) or not (latest lines). 
In the first case, while the two best fit models describe the data with the same goodness ($\Delta\chi^2=0$), the Bayesian analysis shows a \textit{strong} preference of the data for the $\Lambda$CDM+$r$ model. 
This result can be well explained and understood by considering two factors. 
Firstly, the degeneration between $\alpha$ and $H^2$ parameters and the consequent deterioration of the $H^2$ constraint with respect to the standard model primordial amplitude parameter.
Secondly, and more important,  the fact that we are comparing our model (with tensor index free to vary) with a $\Lambda$CDM+$r$ one (which assumes the inflation consistency relation). 
If we relax such relation in the $\Lambda$CDM+$r$+$n_t$ model, we obtain a $n_t$ value of $1.6 \pm 0.9$ and a Bayes factor of 
 $\ln \mathit{B}_{ij} = +1.6$. This implies a \textit{weak} evidence in favor of the SDB model with respect to the $\Lambda$CDM. %with relaxed consistency relation. 
%If, on the other hand, we compare the $\Lambda CDM$ model with and without imposed consistency relation, we obtain a strong evidence in favour of the first ($\ln \mathit{B}_{ij} = +17.3$). 
We also note that, if the tensor index is left free to vary in the standard cosmological model, a significant deterioration in its  Bayesian evidence is produced.
Moreover, as said above, the improved evidence of the SDB model with respect to the latter  is due to our parametrization choice on the tensor amplitude (e.g. in terms of $H$), which leads to the sharp constraints on the parameter $r$.

%comments on Fig. ns_nt
Finally, in order to compare the predictions of the SGC and SDB models, we derive the scalar spectral index parameter $n_{s}$ for the SDB scenario, from the $\alpha$ constraints, using Eq. \ref{pdfns2}.
In Fig. (\ref{fig:ns_nt}), we show the observational bounds on $n_{s}$ and $n_{t}$ plane. 
The red and the blue lines show, respectively, the approximated consistency relations of inflation ($n_t \approx n_s-1$) and SGC ($n_t \approx -(n_s-1)$), where both can be recovered by specific values of the SDB model parameters. As can be seen, both models are compatible with the data at 1$\sigma$, although the SGC (and SDB) combination of parameters lies closer to the mean value. In other words, this amounts to saying that the current CMB data cannot distinguish between standard inflation and String Gas Cosmology (or the specific SDB model associated to it).

\section{Conclusions}
\label{Sec:Conclusions}

The assumption of the breaking of spatial diffeomorphism invariance changes the prediction of simple slow-roll inflationary models, allowing to obtain values of $n_t>0$. Through an observational analysis we have obtained constraints on the parameters of the model  and also tested the specific combinations of values which recover the consistency relation of the SGC. While the scalar predictions of the SDB model are very similar to the ones expected from usual inflation, the tensorial predictions can be quite different. The parameter space of tensor parameters allowed within 1$\sigma$ are mostly dominated by values of the SDB parameter $m_{2}>0$. Also, we have shown that  the parameters of the SDB model, which recover the SGC and inflationary consistency relation, are compatible with the data at 1$\sigma$. 

We have obtained an upper limit on the tensor-to-scalar ratio for the SDB model that is smaller than the one predicted by the usual inflation. Moreover,  through a Bayesian analysis we have shown that, by comparing the SDB model and $\Lambda$CDM+$r$ scenario, which assumes the inflation consistency relation, a strong evidence in favor of the latter is found. On the other hand, when the consistency relation is relaxed we have found a weak evidence in favor of the model with diffeomorphism breaking. Finally, it is worth emphasizing that there are great prospects that the next generation of experiments will improve the results obtained here, which encourages further analyzes of the SDB and SGC models. Furthermore, since the SGC consistency relation lies within the range of allowed values of the SDB  parameters (within 1$\sigma$), we stress the importance of analyzing  other predictions of the model, like the non-Gaussianities for instance \cite{nongauss}.

\section*{Acknowledgements}

We thank Prof. Robert H. Brandenberger for comments on our draft.
LG is supported by Coordena\c{c}\~{a}o  de Aperfei\c{c}oamento de Pessoal
de Nivel Superior (CAPES) (88887.116715/2016-00).
MB acknowledges financial support of the Funda\c{c}\~{a}o Carlos Chagas Filho de Amparo \`{a} Pesquisa do Estado do Rio de Janeiro (FAPERJ - fellowship {\textit{Nota 10}}). 
JSA is supported by Conselho Nacional de Desenvolvimento
Cientifico e Tecnologico (CNPq) and FAPERJ. 
We also acknowledge the authors of the CosmoMC (A. Lewis) and Multinest (F. Feroz) codes.

\end{document}